\documentclass[twocolumn,twocolappendix]{aastex631}
\begin{document}
\title{The Dzhanibekov Effect as a Possible Source of Magnetar Activity}
\author[0000-0001-6010-714X]{Tomoki Wada}
\affiliation{Frontier Research Institute for Interdisciplinary Sciences, Tohoku University, Aoba, Sendai, 980-8578, Japan}
\affiliation{Astronomical Institute, Graduate School of Science, Tohoku University, Aoba, Sendai, 980-8578, Japan}
\affiliation{Institute for Cosmic Ray Research, The University of Tokyo, 5-1-5 Kashiwanoha, Kashiwa, Chiba, 277-8582, Japan}
\author{Jiro Shimoda}
\affiliation{Institute for Cosmic Ray Research, The University of Tokyo, 5-1-5 Kashiwanoha, Kashiwa, Chiba, 277-8582, Japan}
%
\begin{abstract}
Magnetars, which are neutron stars with strong magnetic fields,
exhibit occasional bursting activities.
The shape of a magnetar is not perfectly spherical
due to the Lorentz force exerted by its strong magnetic fields
and is 
described as a triaxial body.
We study the unstable free precession in 
a triaxial magnetar;
one of the principal axes undergoes an upside-down flip.
This flip is known as the
Dzhanibekov effect.
We find that during the flip, the Euler force can suddenly disturb
the force balance on
the surface layer of the magnetar, potentially leading to
plastic flow of the layer.
This, in turn, may trigger different forms of magnetar activity,
such as the emission of the bursts and/or of gravitational waves.
\end{abstract}
\section{Introduction} \label{sec:Int}
Magnetars are neutron stars whose
X-ray luminosity exceeds the spin-down luminosity \citep{DunTho1992}.
Almost all magnetars 
possess
strong magnetic fields of $\sim 10^{14}$--$10^{15}\,{\rm G}$ and
exhibit energetic bursting activities, such as short bursts, outbursts, and
flares \citep[][for reviews]{HarLai2006,TurZan2015,KasBel2017,EnoKis2019_neutronstar}.
These bursts are thought to be triggered by 
transient activities,
such as starquakes, which suddenly release magnetic energy \citep{ThoDun1995,ThoDun1996}.
Magnetars are also one of the origins of
fast radio bursts,
as indicated by
the detection of FRB 20200428A
from the
Galactic magnetar, SGR 1935+2154
\citep{Boc2020,CHIME2020_200428,Mer2020,Li2021,Rid2021,Tav2021}.
Magnetars are also expected to be one of the origins of
ultrahigh-energy cosmic rays \citep[e.g.,][]{Aro2003,AsaYam2006,Kot2011}.
\par
The strong Lorentz force,
counteracting the elastic force, 
can deform the magnetar 
from 
a perfectly spherical shape. 
If a
neutron star is in a binary system,
the accretion from the companion also causes such deformations \citep{FujKis2022}.
%
Her X-1 is an X-ray pulsar
in such an accretion system.
%
In addition to the 1.24-second spin period of the neutron star and the 1.7-day orbital period, a long-term 35-day X-ray flux modulation is observed \citep{TanGur1972}.
This 35-day modulation can be explained by a free precession of a triaxial body
\citep{ShaPos1998,KolSha2022,HeyDor2023}.
The ratio of the spin period to the precession period
of Her X-1 suggests 
its ellipticity 
of $\delta\sim10^{-6}$ \citep{KolSha2022}.
Also, continuous gravitational wave observations constrain the ellipticity of
neutron stars independently as $\delta\lesssim10^{-6}$
at the frequency of $\sim 10^3\,{\rm Hz}$, and $\delta\lesssim1$ at $\sim 10\,{\rm Hz}$ \citep{LIGO2022_cont}.
These observations support the existence of triaxial neutron stars.
\par
The rotation around the second principal axis of a triaxial body can be
unstable \citep[e.g.,][]{AshChi1991,MarGui2020}.
If the rotation is dominant around the second principal axis initially,
the direction of the second principal axis reverses. Hereafter,
this phenomenon is
referred to as a `flip'.
The timescale of the flip can be much shorter than the precession period.
This flip
is the so-called Dzhanibekov effect \citep[e.g.,][for a neutron star]{ShaPos1998}, an unstable case of the free precession, 
whereas the precession explaining the superperiodicity of
Her~X-1 is a stable case \citep{KolSha2022,HeyDor2023}.
The connection between the flip and the magnetar bursts has not been previously discussed,
despite the
flip being a potential origin of
transient events at the neutron star.

In this paper, we study the possible connection between the magnetar bursts and the flips of the triaxial magnetars.
In Sectioin~\ref{sec:Tri},
we review the equation of motion of a triaxial magnetar and demonstrate the occurrence of the Dzhanibekov effect.
%
%
%
%
Section~\ref{sec:Eul} shows that the Euler force on the
magnetar's surface layer
is stronger than the crustal tensile strength during the flip,
and suggests that this force can trigger bursting activities.
%
%
Section~\ref{sec:Con} is devoted to the conclusion.
We use $Q_{,x}=Q/10^x$ in cgs units unless otherwise noted.

\section{Triaxial Free Precession and Flip of a Magnetar}\label{sec:Tri}
\begin{figure}
  \center
  \includegraphics[width=\linewidth]{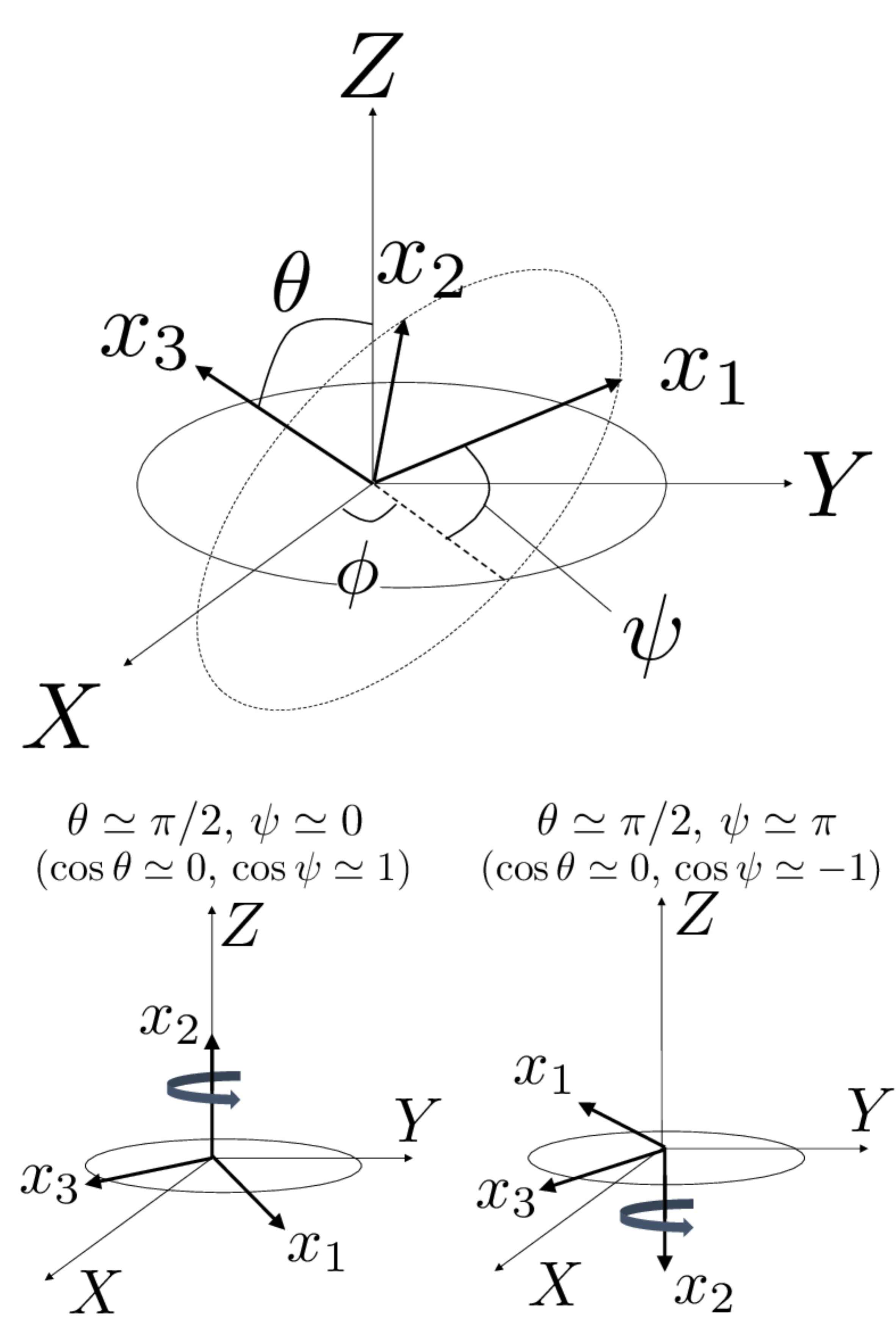}
  \caption{
    Definition of Euler angles (top panel).
    Typical values of the Euler angles that appear in this paper (bottom panels).
    The configuration in the bottom left panel corresponds to $\cos\theta\simeq0$ and $\cos\psi\simeq1$ and that in the bottom right panel corresponds to $\cos\theta\simeq0$ and $\cos\psi\simeq-1$ (see also Figure~\ref{fig:angle}).
    During these configurations, the magnetar mainly rotates around the $x_2$ axis (i.e., $\Omega_2\gg \Omega_1,\Omega_3$)
    although the sign of $\Omega_2$ is different.
    \label{fig:def_angle}
}   
\end{figure}

We consider the motion of a triaxial magnetar \citep[][]{LandauLifshitz1969}.
We assume that the magnetar is triaxial and
that the spin axis does not initially coincide with
any
of the principal axes of inertia.
It is also assumed that the magnetar rotates as a rigid body with a constant moment
of inertia, and 
the interaction between the crust and
the
neutron star interior is neglected \citep[e.g.][]{Lin2007,GogBar2019}.
The three eigenvalues of the moment of inertia tensor, $I_i\,(i=1,2,3)$, are defined as $I_3>I_2>I_1$.
The energy and the angular momentum
are expressed as follows,
\begin{eqnarray}
E&=&\frac{1}{2}\sum_{i=1}^3 I_i\Omega_i{}^2,\label{eq:energy}\\
{\mathcal M}^2&=&\sum_{i=1}^3 I_i{}^2\Omega_i{}^2,\label{eq:momentum}
\end{eqnarray}
where $E$ and ${\mathcal M}$ are the kinetic energy of the rotation and
the total angular momentum of the magnetar, respectively.
$\Omega_i~({\rm i=1,2,3})$ are
the components of the angular velocity vector, $\vec{\Omega}$, along
each principal axis of inertia, $x_1$, $x_2$, and $x_3$ (see top panel of Figure~\ref{fig:def_angle}).
If the magnetar, 
with a mass of $M_{\rm NS}$ and 
a radius of $R_{\rm NS}$, is a uniform sphere,
%
%
then the eigenvalues of
the moment of inertia tensor
satisfy $I_1=I_2=I_3=(2/5)M_{\rm NS}R_{\rm NS}{}^2$.
%
In this paper, we set 
$I_1$ and $I_3$ to be slightly different 
from $(2/5)M_{\rm NS}R_{\rm NS}{}^2$ as $I_3>I_2>I_1$.
The differences 
between $I_1, \, I_2$, and $I_3$ will be discussed below.

We use Euler angles to describe the rotational motion of the magnetar.
The top panel of Figure~\ref{fig:def_angle} shows the definitions of the Euler angles, $\theta,\,\psi$, and $\phi$.
The inertial frame is denoted by $(X,Y,Z)$.
The axes $(x_1,\,x_2,\,x_3)$
represent the body axes.
$\phi$ is
the angle
between the $X$ axis and $\vec{e}_Z\times\vec{e}_{x_3}$,
where $\vec{e}_Z$ and $\vec{e}_{x_3}$ are the unit vectors in the $Z$ and $x_3$ directions,
respectively.
$\psi$ is the
angle
between the $x_1$ axis and $\vec{e}_Z\times\vec{e}_{x_3}$.
$\theta$ is the angle between the $Z$ axis and the $x_3$ axis.
For $\theta=0, \,\psi=0,$ and $\phi=0$, the $x_1$, $x_2$, and $x_3$ axes respectively coincide with the $X,\,Y,$ and $Z$ axes.

The equations of motion (Euler's equations for free rotation) are 
\begin{eqnarray}
  I_1\dot{\Omega}_1&=&(I_2-I_3)\Omega_2\Omega_3\label{eq:eom1},\\
  I_2\dot{\Omega}_2&=&(I_3-I_1)\Omega_3\Omega_1,\\
  I_3\dot{\Omega}_3&=&(I_1-I_2)\Omega_1\Omega_2\label{eq:eom3},
\end{eqnarray}
where a dot denotes
a derivative of a function with respect to the time, $t$.
For simplicity, we assume that $(I_3-I_2)/I_2=(I_2-I_1)/I_2$ and set $I_2=(2/5)M_{\rm NS}R_{\rm NS}^2,\,I_1=I_2(1-\delta), \, I_3=I_2(1+\delta)$, where $\delta$ is a dimensionless parameter describing the non-sphericity.
$\delta$ is much smaller than unity and may be $\delta\sim 10^{-6}$-- $10^{-11}$
(\citealp{KolSha2022} for the observational values and \citealp{ZanLai2020} for a summary
of the theoretical values).
There are 
several possible origins of the non-sphericity: the internal magnetic field of the
magnetar, the elasticity of the magnetar's surface layer, the magnetic field exterior to the magnetar,
the effect of magnetosphere plasma, and so on \citep[see][for summary]{ZanLai2020}.

In this 
study, we focus on the case of $\Omega_{2,\rm ini}\gg\Omega_{1,\rm ini},\Omega_{3,\rm ini}$, where $\Omega_{i,\rm ini}$ ($i=1,2,3$) are the initial values of $\Omega_i$.
The equation of motion for $\Omega_1$ under this assumption is rewritten as
\begin{eqnarray}
  \ddot{\Omega}_1\simeq \frac{(I_3-I_2)(I_2-I_1)}{I_1 I_3}\Omega_2^2 \Omega_1
  \simeq \delta^2\Omega_2^2 \Omega_1,
\end{eqnarray}
indicating an exponential growth of $\Omega_1$.
The equation of motion for $\Omega_3$ is similar to that for $\Omega_1$.
The timescale of this instability is 
\begin{eqnarray}
  T_{\rm ins}\simeq \sqrt{\frac{I_1I_3}{(I_3-I_2)(I_2-I_1)\Omega_2^2}}\sim \delta^{-1}\frac{P}{2\pi},
  \label{eq:timescale}
\end{eqnarray}
where $P$ is the spin period, and we used $\Omega_2\sim (2\pi)/P$.
According to the detailed analysis in Appendix~\ref{sec:appendix}, $\Omega_i\,(i=1,2,3)$
are periodic functions of time.
The Euler angles $\theta$ and $\psi$ are also periodic functions of time,
with a period of
\begin{eqnarray}
  T&\sim&4K\delta^{-1}\Omega_{2,\rm ini}{}^{-1}
  \sim 200\, {\rm yr}\,K_{1}\delta_{-9}{}^{-1}P_{,0},
  \label{eq:period}
\end{eqnarray}
where $K$ is the complete elliptic integral of the first kind, as given in Appendix \ref{sec:appendix}.

\begin{figure}
  \center
  \includegraphics[width=\linewidth]{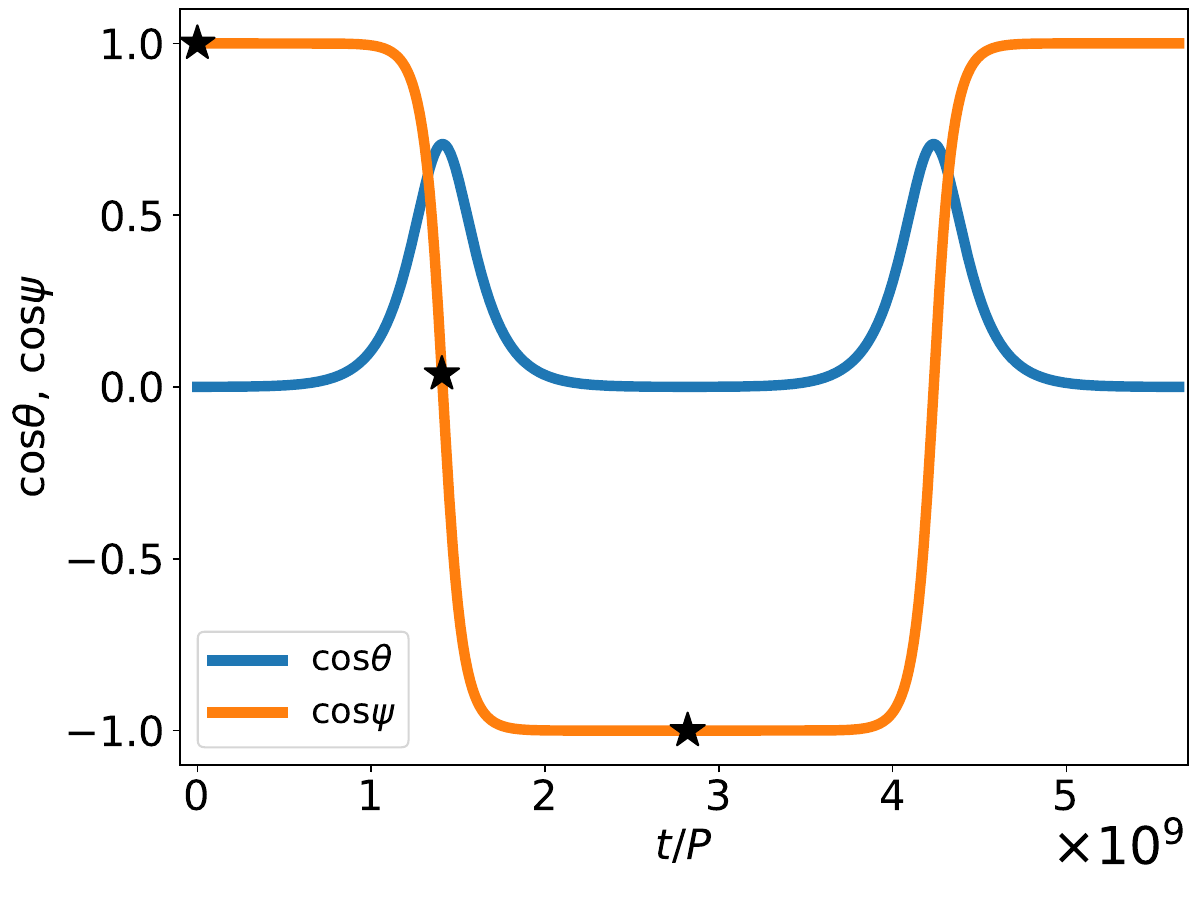}
  \caption{
    Time evolution of the Euler angles, $\theta,\psi$, during one period $T$.
    The $x_2$ axis flips ($\cos\psi$ suddenly changes from $1$ to $-1$, or from $-1$ to $1$) around $t\sim1.4\times 10^9 P$ and $t\sim4.2\times 10^9P$.
    The configurations before and after the flip are shown in the bottom panels of Figure~\ref{fig:def_angle}.
    \label{fig:angle}
  }
\end{figure}

Figure~\ref{fig:angle} shows the time dependences of the Euler angles, where
we adopt $\delta=10^{-9}$ and set the initial angular velocities
as $\,\Omega_{1,\rm ini}=\Omega_{3,\rm ini}=0.1(2\pi/P),\,\Omega_{2,\rm ini}=\sqrt{1-2\times0.1^2}(2\pi/P)$.
For these parameters, $K$ equals $8.8$, and the
period 
is $T\sim 5.7\times10^9P$.
The $x_2$ axis flips from $\psi=0$ to $\psi=\pi$, or from $\psi=\pi$ to $\psi=0$ during a relatively short interval of
$T$ (see also the bottom of Figure~\ref{fig:def_angle}),
known as the ``Dzhanibekov effect'', in which the initial rotation around the second principal axis
is dominant.
To validate
the free precession approximation, $T$ should be smaller than the spin-down timescale.
The spin-down timescale of the neutron star is \citep[e.g.,][]{ShapiroTeukolsky1983}
\begin{eqnarray}
  t_{\rm sd}\sim \frac{3I_2c^3}{2B_{\rm NS}{}^2R_{\rm NS}{}^6\Omega^2}.
\end{eqnarray}
The condition $t_{\rm sd}\gtrsim T$ 
provides the lower limit on the spin period as,
\begin{eqnarray}
      P
      \sim3\times10^{-1}{\rm s}\,K_1\delta_{-9}{}^{-1}B_{\rm NS,15}{}^{2}R_{\rm NS,6}{}^4
    M_{\rm NS,\odot}{}^{-1},
    \label{eq:sd_limit}
\end{eqnarray}
where $M_{\rm NS,\odot}=M_{\rm NS}/M_{\odot}$, and $M_{\odot}$ is the solar mass.
This lower limit is shorter than the spin period of 
known magnetars, $\sim 1\,{\rm s}$.


The estimated prevalence of magnetars exhibiting the Dzhanibekov effect
is as follows.
Primarily, the magnetar must be triaxial.
This condition would be satisfied because 
no star is perfectly spherical or axisymmetric.
Secondly, the spin around the second principal axis, $\Omega_2$,
must be the dominant component of $\vec{\Omega}$.
In an assumed uniform, probabilistic selection of the spin axis,
approximately one-third of magnetars satisfy this condition.
This assumption may be optimistic because the initial angular
momentum can be redistributed by viscosity \citep[e.g.,][]{DalSte2018}.
Such a redistribution may reduce the prevalence of magnetars in the unstable configuration,
although the stability of the spin
remains a matter of debate \citep{Lin2007}.
Thirdly, the total period of the flip, $T$, must be smaller than the observation time.
Assuming an observation
term of $\sim$ 100 years,\footnote{
The spin-period evolutions of magnetars are observed with high precisions, 
and the bursting activity 
associated with the flip has not been identified in the last $\sim 50$ years.
So,
we assume a long observation time here.
} 
this condition, $T\lesssim 100\,{\rm yr}$, 
provides the lower limit on the ellipticity
as $\delta\gtrsim2\times10^{-9}$
(see Equation~\ref{eq:period}).
Thus, the expected parameter
of our model is $\delta\sim {\mathcal O}(10^{-9})$.
Note that a smaller $\delta$ may be preferred to explain the stable pulse period of magnetars.
%
%
As will be discussed in Section~\ref{sec:Eul}, the flip may result in a bursting activity. The event rates of associating 
bursts may provide constraints on the actual value of $\delta$. For example, if the flip 
triggers the giant flares, the small $\delta\sim 10^{-9}$ should be realized to explain the event rate of the flares. However, since the details of associated flares (such as lightcurves and spectra) are beyond the scope of this paper, no comparison is made, which should be addressed in future work.
%
%
\section{Euler Force and Possible Bursting Activities during The Flip}\label{sec:Eul}

\begin{figure*}[htb]
  \begin{center}
    \begin{tabular}{c}
      \begin{minipage}{0.33\hsize}
        \begin{center}
          \includegraphics[clip,width=\textwidth]{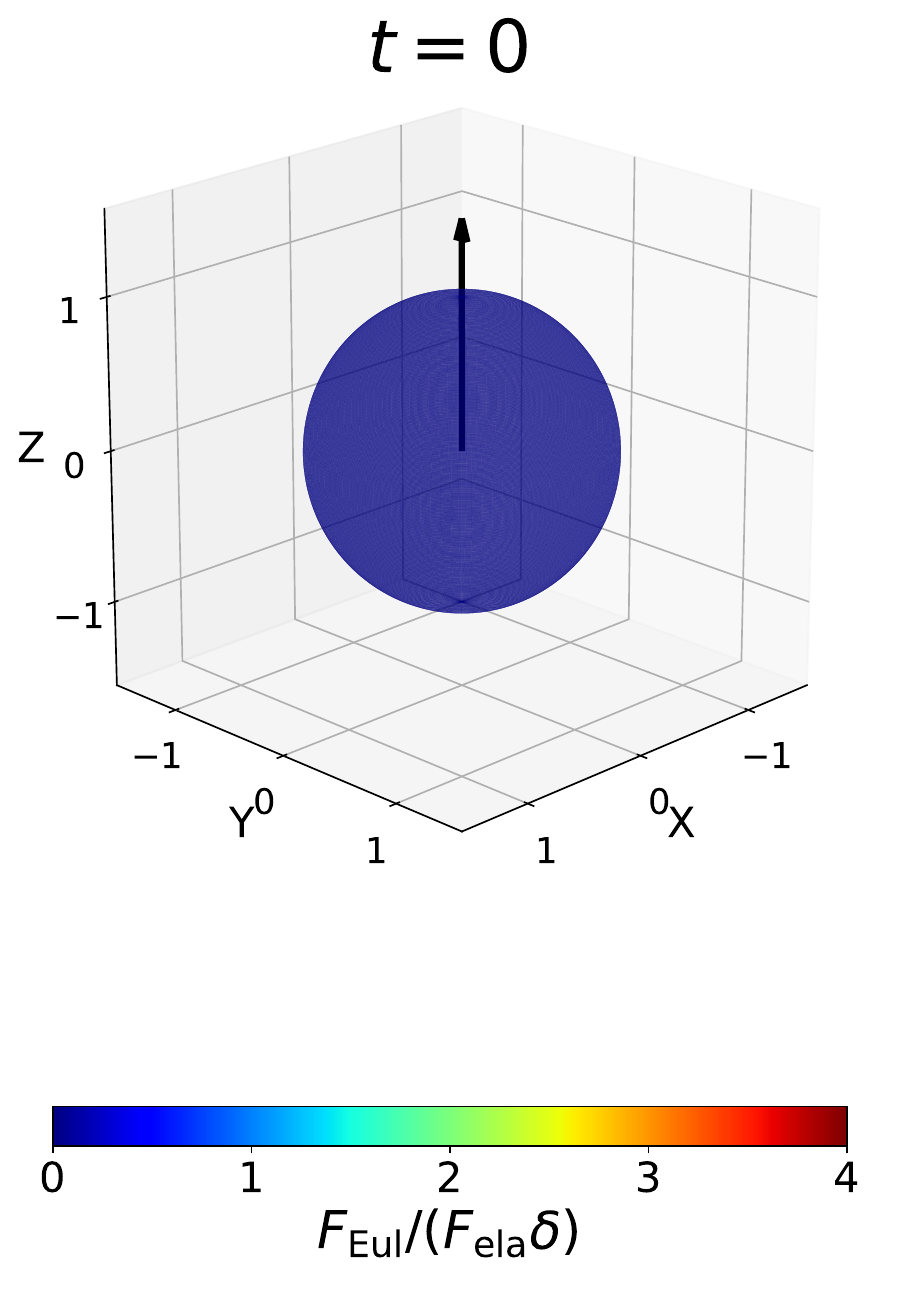}
        \end{center}
      \end{minipage}
      \begin{minipage}{0.33\hsize}
        \begin{center}
          \includegraphics[width=\textwidth]{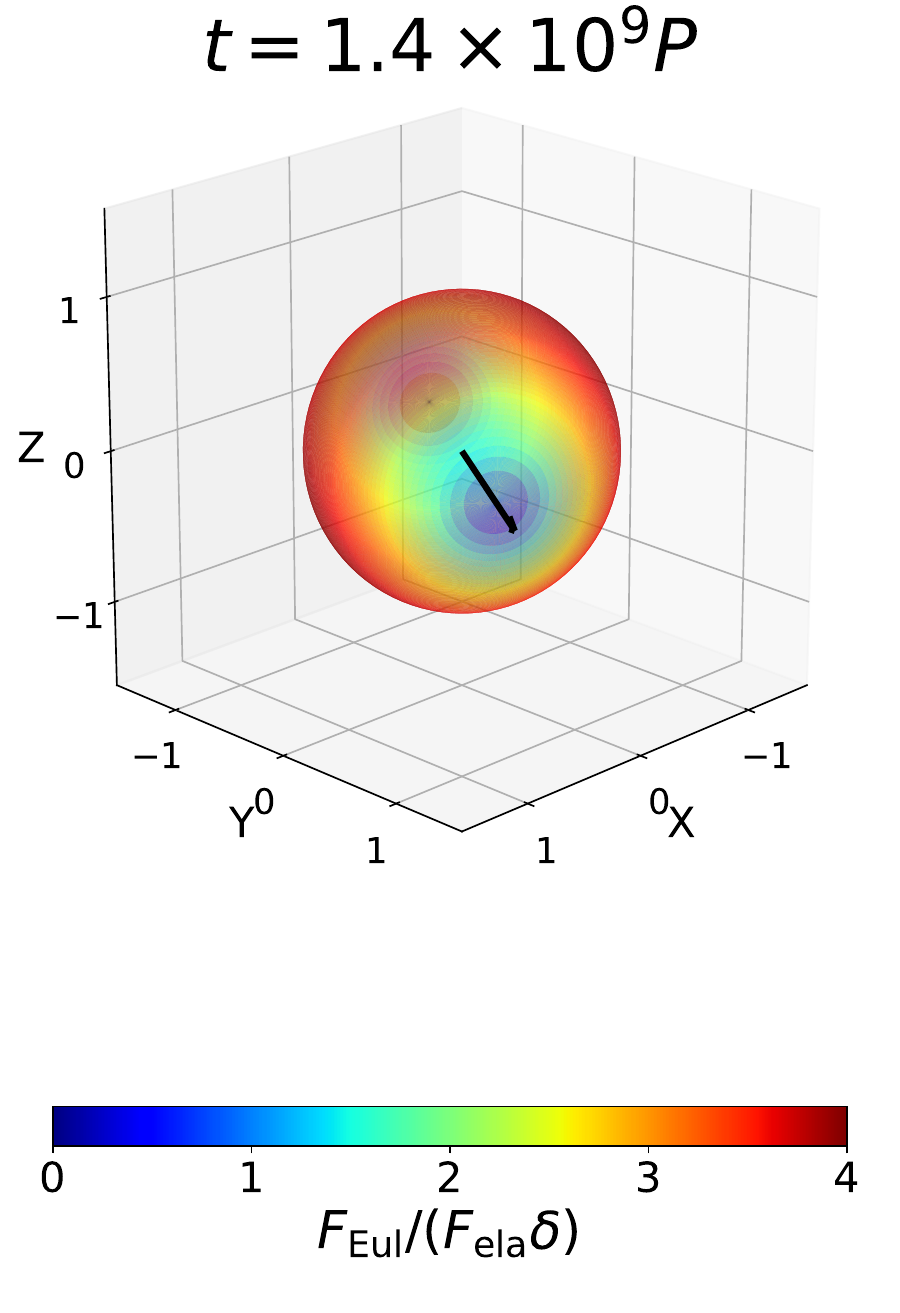}
        \end{center}
      \end{minipage}
              \begin{minipage}{0.33\hsize}
        \begin{center}
          \includegraphics[width=\textwidth]{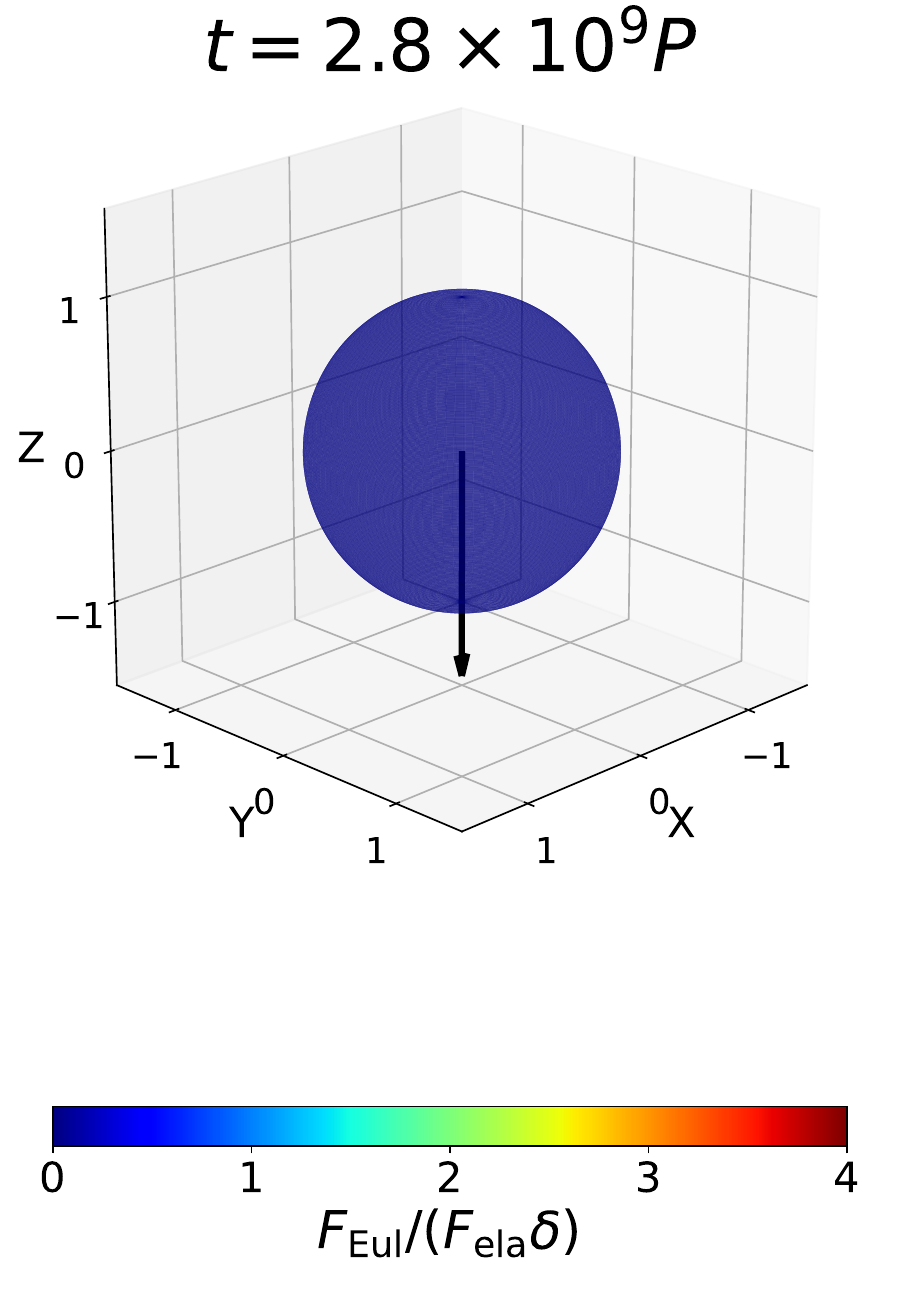}
        \end{center}
      \end{minipage}
    \end{tabular}
    \caption{
    Magnitude of the Euler force on the crust, $F_{\rm Eul}$, normalized by $\delta F_{\rm Ela}$ (see Equation~\ref{eq:dela}).
    The left panel shows the force at $t=0$ (before the flip), the middle panel shows it at $t=1.4\times 10^9P$ (during the flip), and the right panel shows it at $t=2.8\times 10^9P$ (after the flip) (see also Figures~\ref{fig:def_angle} and \ref{fig:angle}).
    The black allow shows the $x_2$ axis.
    During the flip of the $x_2$ axis (middle panel), the Euler force suddenly becomes stronger than before and after the flip.
            }
    \label{fig:euler}
  \end{center}
\end{figure*}

\begin{figure}
  \center
  \includegraphics[width=\linewidth]{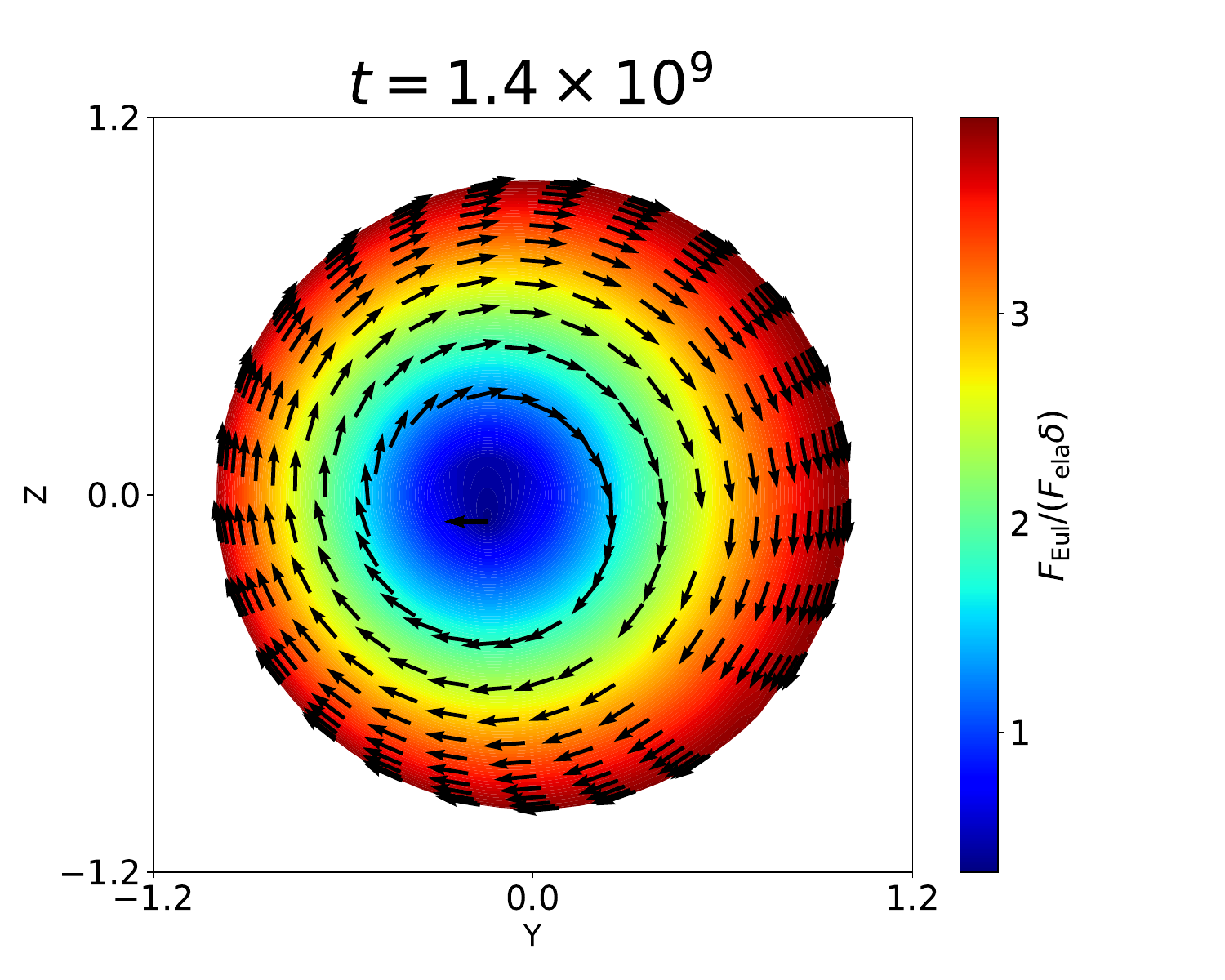}
  \caption{
    Direction and strength of the Euler force seen from a positive $X$.
    The color contour shows the strength of the Euler force normalized by $\delta F_{\rm ela}$, and the arrows show the direction of the Euler forcs on the
    magnetar's surface.
    \label{fig:euler_vec}
  }
\end{figure}

During the 
rotation, the magnetar
is subjected to
the effects of the inertia force.
In the following, we consider the outer layer of the magnetar,
which is commonly referred to as the crust \citep[e.g.,][]{ChaHae2008}.
The inertia force is composed of three components: the Euler force, the Coriolis force, and the centrifugal force \citep[e.g.,][]{LandauLifshitz1969}.
The Euler force per unit mass is $\vec{r}\times\dot{\vec{\Omega}}$, where $\vec{r}$ is 
the position vector in the rotating frame.
The Coriolis force per unit mass is $2\dot{\vec{r}}\times\vec{\Omega}$, and the centrifugal force per unit mass is $\vec{\Omega}\times(\vec{r}\times\vec{\Omega})$.
The Coriolis force
is negligible due to the significantly longer timescale associated with
the motion of the crust in comparison to the other relevant timescales
(i.e., $|\dot{\vec{r}}|/|\vec{r}| \ll |\dot{\vec{\Omega}}|/|\vec{\Omega}|, |\vec{\Omega}|$).
The force balance among the centrifugal force, the pressure gradient, the elastic force, and the Lorentz force on the crust determines the shape of the magnetar.
The Euler force appears only during the flip as an additional force
because of its proportionality to $\dot{\vec{\Omega}}$.
We evaluate the Euler force on the crust
below and conclude
that the crust can plastically flow during the flip.

Figure~\ref{fig:euler} shows the temporal evolution of the Euler force on
the crust at three distinct time points: prior to, during, and post the flip.
During the flip, 
the Euler force exhibits a sudden increase in strength
relative to the pre-flip state.
After the flip, the Euler force returns to its pre-flip strength.
The magnitude of the Euler force per unit volume element of the crust during the flip is
\begin{eqnarray}
  F_{\rm Eul}\sim \rho R_{\rm NS}\dot{\Omega}\sim \rho R_{\rm NS}\Omega^2\delta,
  \label{eq:euler}
\end{eqnarray}
where we use Equations~(\ref{eq:eom1})--(\ref{eq:eom3}).

The force balance on the crust is suddenly broken by the Euler force that is suddenly applied.
Prior to the flip, the Lorentz force, $F_{\rm Lor}$, and the elastic force, $F_{\rm ela}$, would be 
balanced\footnote{
Strictly speaking, the force balanced by the elastic force does not necessarily
be the Lorentz force.}\citep[e.g.,][]{ChaHae2008,KojKis2021}.

If the force balance between the Lorentz force and the elastic force determines
the triaxial shape,
which is characterized by $\delta$,  we may obtain
\begin{eqnarray}
  |F_{\rm Lor}-F_{\rm ela}|&\sim& \delta F_{\rm ela}\sim \delta\frac{\mu\sigma^2}{\ell},
  \label{eq:dela}
\end{eqnarray}
where $\mu$ and $\sigma$ are shear modulus and shear-strain of the crust \citep[e.g.,][]{KojKis2021}.
$\ell$ is the minimal length scale of the plastic flow that we will estimate in the following.
When the Euler force becomes as strong as  $|F_{\rm Lor}-F_{\rm ela}|_{\sigma=\sigma_c}$, the magnetar crust will plastically flow.
Using Equations~(\ref{eq:euler}) and (\ref{eq:dela}), the condition for the plastic flow is estimated as
\begin{eqnarray}
  \frac{F_{\rm Eul}}{|F_{\rm Lor}-F_{\rm ela}|}
    &\sim&4\,\ell_{3}R_{\rm NS,6}P_{,0}{}^{-2}\sigma_{c,-2}{}^{-2}\left(\frac{\mu}{\rho}\right)^{-1}_{14}>1,\nonumber\\
  \label{eq:balance}
\end{eqnarray}
where $\sigma_c$ is the critical shear-strain when the crust plastically flows, and we adopt $\sigma_c\sim 10^{-2}$ and $\mu/\rho\sim 10^{14}\,{\rm cm^2s^{-2}}(A/56)^{-4/3}(Z/26)^{2}\rho_{,4}{}^{1/3}$ \citep{OgaIch1990,ChaHae2008}.
The ratio of the shear modulus to the mass density, $\mu/\rho$, is 
calculated for temperatures up to the melting conditions, including
the effects of thermal fluctuations, under the assumption that
the magnetar crust element is an isotropic body-centered cubic polycrystal.
The maximum shear-strain, $\sigma_c$, is not well known and varies from $\sim10^{-5}$ to $\sim10^{-1}$ depending on the approach \citep[summarized in][]{KojKis2021}.
In this paper, we adopt $\sigma_c\sim 10^{-2}$.
Equation~(\ref{eq:balance}) 
indicates that the Euler force can 
disrupt the force balance between the Lorentz force and the critical elastic force if $\ell$ is larger than $\sim 10^3\,{\rm cm}$ for $\sigma_c=10^{-2}$.

Figure~\ref{fig:euler_vec} shows the Euler force at
the surface layer of the magnetar.
The Euler force is almost uniform in direction but strongly depends on the position of the crust.
The Euler force, $\vec{r}\times\dot{\vec{\Omega}}$, is zero at
the
two points where $\vec{r}$ is
parallel 
or anti-parallel to $\dot{\vec{\Omega}}$.
In the case of Figure~\ref{fig:euler_vec},
the position vector $\vec{r}$
is parallel to $\dot{\vec{\Omega}}$
in the vicinity of the origin of the $Y$-$Z$ plane.
Conversely, the Euler force is at its maximum where $\vec{r}$ is perpendicular to $\dot{\vec{\Omega}}$: the red ring region in Figure~\ref{fig:euler_vec} and the middle panel of Figure~\ref{fig:euler}.
We expect that the force balance is disturbed 
by this position-dependent Euler force,
resulting in the crust's plastic flow.
The radial component of the Euler force
is zero 
since $\vec{F}_{\rm Eul}=\vec{r}\times \dot{\vec{\Omega}}$ is perpendicular to $\vec{r}$.
We note that $\ell$ is the lower limit of the length scale at which the Euler force can disturb the force balance on the crust, and the length scale of the plastic flow can be larger than $\ell$.
Thus, the flip results in the plastic flow 
of the surface layers with the length scale of $\ell\gtrsim 10^3\,{\rm cm}$.

During the flip, the strong Euler force on the crust may trigger transient activities.
If some energy is released and 
results in forming a fireball,
X-ray photons are emitted from it \citep{ThoDun1995}.
If 
bursts occur in the early and late phases of the flip, polarization signature 
or X-ray spectrum may change \citep{Lyu2002,YanZha2015,TavTur2017,YamLyu2020}.
This is because 
they depend on the configuration of the magnetic field, 
which changes before and after the flip
for the observer.
If the fireball is formed near the magnetic pole, it expands and accompanies a relativistic outflow \citep{ThoDun2001,Iok2020,YanZha2021,WadIok2023}.
Due to the strong Euler force, many baryons might be loaded on the fireball, 
forming a relativistic outflow with high kinetic luminosity \citep{WadIok2023,WadAsa2024}.
\par
Since the Dzhanibekov effect may be a rare event, as mentioned in
Section~\ref{sec:Tri}, this scenario may not
fully account for all observed
bursting activities of magnetars. 
We will construct a more detailed model to study a proper prediction of this scenario in future work.
In addition to
X-ray transients,
the gravitational waves from triaxial neutron stars may also be expected as studied by \cite{Zim1980}.
%

Some studies discuss a correlation between transient activities
and flips in neutron stars.
Flip-like phenomena can occur in newborn neutron stars due to their internal bulk viscosity in the hydrodynamic regime \citep{DalSte2018}.
This results in the emission of transient gravitational waves,
and potentially the energy injection into short gamma-ray bursts.
Such flips in newborn neutron stars
may 
be related
to the initial distribution of the spin axis in our scenario.
It is pointed out that
starquake-like phenomena are
caused by the angular momentum transfer
between the crust and core \citep[e.g.,][]{Lin2007,GogBar2019}.
The rigid body approximation for magnetars may be invalidated by an efficient transfer
of the internal angular momentum, which is also responsible for the evolution of the spin.
Even if a magnetar 
behaves as a rigid body, the rotation around the second principal axis should be dominant to satisfy the unstable condition.
\cite{KolSha2022} studies 35-days-periodic modulation of Her X-1 adopting the ``stable" free precessions,
where ``stable" means
the absence of the flips.
Our model covers the unstable" parameter space that has not previously been studied in detail and is the necessary piece for studying magnetars from the cradle to the grave.

\section{Conclusion}\label{sec:Con}
We investigate the free precession of a triaxial magnetar and the possible astrophysical transients associated with the precession.
We show that a triaxial magnetar can flip under the initial condition that the spin around the second principal axis is dominant (Section~\ref{sec:Tri}).
From the time evolution of the angular velocities, we evaluate the inertial force on the magnetar crust and show that the Euler force is strongest during the flip.
The Euler force can be strong enough to 
disturb the force balance on the crust.
If the 
magnetar surface layer plastically flows 
due to the Euler force, astrophysical bursting activities may be triggered, such as X-ray transients (Section~\ref{sec:Eul}).

\section*{Acknowledgement}
We thank K. Asano, R. Goto, Y. Ichinose, T. Kawashima, Y. Kusafuka, K. Nishiwaki for daily discussions.
We also thank the anonymous referee for helpful comments and suggestions.
This work is supported by Grants-in-Aid for Scientific Research No.	22K20366 (TW) from the Ministry of Education, Culture, Sports, Science and Technology (MEXT) of Japan.
Our study is also supported by the joint research program of the Institute for Cosmic Ray Research (ICRR), the University of Tokyo.


\bibliography{cite}{}
\bibliographystyle{aasjournal}

\appendix
\section{Analytic solution}\label{sec:appendix}
In this section, we review the analytic solution of the Euler equations (Equations \ref{eq:eom1}--\ref{eq:eom3}).
The analytic solutions of Equations (\ref{eq:eom1})--(\ref{eq:eom3}) for $\mathcal M^2>2EI_2$ are \citep{LandauLifshitz1969}
\begin{eqnarray}
  \Omega_1&=&\sqrt{\frac{2EI_3-{\mathcal M}^2}{I_1(I_3-I_1)}}{\rm cn}(\tau;\,k),\label{eq:sol1}\\
  \Omega_2&=&\sqrt{\frac{2EI_3-{\mathcal M}^2}{I_2(I_3-I_2)}}{\rm sn}(\tau;\,k),\\
  \Omega_3&=&\sqrt{\frac{{\mathcal M}^2-2EI_1}{I_3(I_3-I_1)}}{\rm dn}(\tau;\,k)\label{eq:sol3}
\end{eqnarray}
where
\begin{eqnarray}
  \tau&=&t\sqrt{\frac{(I_3-I_2)({\mathcal M}^2-2EI_1)}{I_1I_2I_3}},\\
  k^2&=&\frac{(I_2-I_1)(2EI_3-{\mathcal M}^2)}{(I_3-I_2)({\mathcal M}^2-2EI_1)},
\end{eqnarray}
and ${\rm sn}(x)$, ${\rm cn}(x)$, and ${\rm dn}(x)$ are the Jacobi elliptic functions.
The Jacobi elliptic functions are periodic with respect to the real argument $t$ and its period is
\begin{eqnarray}
  T&=&4K\sqrt{\frac{I_1I_2I_3}{(I_3-I_2)({\mathcal M}^2-2EI_1)}}
\end{eqnarray}
where
\begin{eqnarray}
  K&=&\int_0^1\frac{ds}{\sqrt{(1-s^2)(1-k^2s^2)}}
\end{eqnarray}
If we set $I_1=I_2(1-\delta), \, I_3=I_2(1+\delta)$, and $\Omega_{2,\rm ini}\gg\Omega_{1,\rm ini},\Omega_{3,\rm ini}$, $k^2$ is expressed as 
\begin{eqnarray}
k^2&=&\frac{1+2(\Omega_{1,\rm ini}/\Omega_{2,\rm ini})^2}{1+2(\Omega_{3,\rm ini}/\Omega_{2,\rm ini})^2}\nonumber\\
  &\times&\left[1-\left(\frac{2(\Omega_{1,\rm ini}/\Omega_{2,\rm ini})^2}{1+2(\Omega_{1,\rm ini}/\Omega_{2,\rm ini})^2}+\frac{2(\Omega_{3,\rm ini}/\Omega_{2,\rm ini})^2}{1+2(\Omega_{3,\rm ini}/\Omega_{2,\rm ini})^2}\right)\delta\right],\nonumber\\
\end{eqnarray}
up to the first order of $\delta$, and the period is
\begin{eqnarray}
  T&\sim&4K\delta^{-1}\Omega_{2,\rm ini}^{-1}
  \sim 0.2\, {\rm yr}\,K_{1}\delta_{-6}^{-1}P_{0}.
\end{eqnarray}
$K$ is a monotonously increasing function for $k$, $K$ equals $\pi/2$ for $k=0$, $K\simeq 8.3$ for $k^2=1-10^{-6}$, and $K\simeq 13$ for $k^2=1-10^{-10}$.
We adopt $K=10$ in this paper.



\end{document}